# GENERATIVE ADVERSARIAL NETWORKS IN FINANCE: AN OVERVIEW

**Florian Eckerli** *
School of Engineering
Zurich University of Applied Sciences
Winterthur, Switzerland
e.florian@hotmail.com

**Joerg Osterrieder***
School of Engineering
Zurich University of Applied Sciences
Winterthur, Switzerland
joerg.osterrieder@zhaw.ch

The Hightech Business and Entrepreneurship Group
Faculty of Behavioural, Management and Social Sciences
University of Twente
Enschede, Netherlands
joerg.osterrieder@utwente.nl

June 11, 2021

## ABSTRACT

Modelling in finance is a challenging task: the data often has complex statistical properties and its inner workings are largely unknown. Deep learning algorithms are making progress in the field of data-driven modelling, but the lack of sufficient data to train these models is currently holding back several new applications. Generative Adversarial Networks (GANs) are a neural network architecture family that has achieved good results in image generation and is being successfully applied to generate time series and other types of financial data. The purpose of this study is to present an overview of how these GANs work, their capabilities and limitations in the current state of research with financial data and present some practical applications in the industry. As a proof of concept, three known GAN architectures were tested on financial time series, and the generated data was evaluated on its statistical properties, yielding solid results. Finally, it was shown that GANs have made considerable progress in their finance applications and can be a solid additional tool for data scientists in this field.

## 1 Introduction to the modelling problem in finance

The finance industry is one of the most influential fields impacted by new developments in AI. More specifically machine learning has been deployed to forecasting, customer service, risk management and portfolio management, among others. The finance industry generates huge amounts of data which offers new opportunities for models to be applied and used to obtain better knowledge and spot opportunities. It also has the resources to apply into research of the most novel techniques.

Financial markets, like the economy, are highly complex systems where it is often impossible to explain macro phenomena by a simple summation of micro processes or events. It's hard to predict the results of actions in the system

*This article is based upon work from COST Action 19130 Fintech and Artificial Intelligence in Finance, supported by COST (European Cooperation in Science and Technology), www.cost.eu. (Action Chair: Joerg Osterrieder)

Financial support by the Swiss National Science Foundation within the project "Mathematics and Fintech - the next revolution in the digital transformation of the Finance industry" is gratefully acknowledged by the corresponding author. This research has also received funding from the European Union's Horizon 2020 research and innovation program FIN-TECH: A Financial supervision and Technology compliance training programme under the grant agreement No 825215 (Topic: ICT-35-2018, Type of action: CSA) and sometimes impossible to find the causes of large abnormalities, or even to single out factors that influenced an event. In the 2010 flash crash[1] for instance, plausible explanations only surfaced years after the episode, and even in specific segments, studies[2] have found that tools for market analysis show a incomplete picture. Modelling in this setting involves dealing with a great deal of uncertainty.

While there are large amounts of data generated by the banking and financial sectors, it's possible to notice the shortage of publicly available datasets originating from these environments. Daily stock price data, for instance, is openly available, but many other types of data are protected by regulations and privacy laws. This scarcity of historical and other types of data has been a limiting barrier to achieve better models. The data-intensive nature of ML models generates the situation where a lack of data limits the potential of data-driven methods.

This paper aims to show an overview of the novel framework for generating synthetic data called Generative Adversarial Networks, also known as GANs, a family of generative models that is being used to produce diverse types of synthetic financial data. GANs are being deployed to solve some of the scarcity of real data, optimize portfolios and trading strategies, among other uses. The goal is to provide a basic understanding of how a GAN works, present the current state of research and current practical applications. Additionally, a proof of concept model was built to generate some synthetic financial time series with a data-driven approach.

## 1.1 Stylized facts of financial time series

The dynamics of financial markets can be defined as examples of complex phenomena[3], since the underlying mechanics of the processes are largely unknown. However, years of research in financial time series have shown that although they may look completely random, variations of asset prices share several significant statistical properties, which are common across a multitude of markets and timeframes. These properties are known as empirical facts, or just stylized facts.

Stylized facts were gathered by taking statistical features of asset returns found in studies about many markets and instruments. A causal analysis of distinct economic scenarios over the globe may expect that being influenced by different events and environment would result in different statistical properties of the said returns. Still, a result of decades of empirical studies of financial markets has shown that some key characteristics are constantly present across markets with different assumed characteristics. Due to these generalizations, the obtained features end up losing in precision, giving the stylized facts a more qualitative aspect. Is important to note that these qualitative properties are not easy to reproduce via modelling of stochastic processes and end up being essential in modelling asset price dynamics, as it is expected that these models can capture/replicate these statistical features.

From the proposed stylized facts[4] a selection of five is portrayed here based on their relevance and use in the modelling literature:

• **Linear unpredictability** or **absence of autocorrelations**: Except for short intra-day time scales, it is expected that asset returns show minimal linear autocorrelations.

• **Fat-tailed distribution** or **heavy tails**: the distribution of returns seems to display a power-law or Pareto-like tail, which in practical terms means that the probability of extreme values occurring is much larger than in a normally distributed dataset.

• **Volatility clustering**: different measures of volatility display a positive autocorrelation over several days. Meaning that there is a tendency of high volatility events to cluster together. Simplifying, large price changes tend to be followed by large changes and small price changes tend to be followed by small changes.

• **Gain/loss asymmetry**: Observations have shown that large plunges in stock prices and index values do no share equally large upward movements.

• **Aggregational Gaussitanity**: Over larger time scales, the distribution of returns looks like a normal distribution.

## 1.2 Models used in finance

Presently there are a few approaches to financial time series modelling, notably there are the **ARCH/GARCH** family of models, which rely on classical statistics and model the change in variance over time in a time series by describing the variance of the current error term as a function of past errors. Often the variance is related to the squares of the previous innovations. As AR (auto regressive) models, they heavily rely on past information to construct a prediction. They are commonly used in time series modelling where time-varying volatility and volatility clustering are present. They don't have a stochastic component since volatility is completely dependent on previous values. ARCH/GARCH models are used to predict the variance at future time steps.

Another used family of models are the Agent-based models (**ABM**s). In agent-based models, the agents are entities, typically represented computationally as objects. These agents hold a state, which can be any data that describes the agent. ABMs simulate interactions of multiple agents to re-create and predict the behaviour of complex phenomena, where the goal is to generate higher level properties from the interaction of lower level agents.

Modelling financial time series is a big challenge, since the dynamics of financial markets are very complex in nature, and the mechanism generating the data, and thus its original distribution, is unknown. Adopting a purely data-driven modelling approach to this problem could provide new solutions or alternative paths by removing a source of bias from modelling.

**Artificial intelligence in finance**

Making use of newly available modelling techniques and processing power, financial analysis has been evolving constantly[5] into more complex modelling methods, such as the deep learning approaches and most recently, Generative Adversarial Networks. GANs may provide good results, since they can generate data by sampling only from real data, often with no additional assumptions or inputs. Avoiding assumptions may be a very important aspect to this type of modelling due to the largely empirical aspect of financial data, the avoidance of possible human bias being infiltrated into the modelling process could possibly be a step forward in this field of study.

## 2   Generative Adversarial Networks

Generative adversarial networks, GANs, are a series of generative machine learning frameworks first introduced by Ian Goodfellow and his collaborators in 2014[6]. They gained a lot of attention due to its simplicity and effectiveness. In a short time span, considerable progress was made to the initial application of GANs - image generation, but also hundreds of different types of GANs were created to optimize for various tasks, from computer vision to fraud detection in banks, the framework presented in 2014 has come a long way. From getting good images on the MNIST dataset and recognizable human faces in the original paper, as of 2021, GANs can generate near perfect human faces with StyleGAN[7], as seen in figure 1, and have a constantly expanding portfolio of applications.

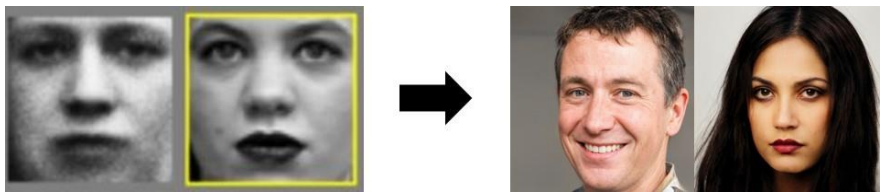

Figure 1: 2014: first generated faces[6] vs. 2020: current state of the art[7]

There are also several applications to finance and financial time series analysis, although compared to other fields, these frameworks are still considered novelties, which makes for an exciting area of research. Since the research is still in its infancy, its reasonable to assume that further applications are yet to be developed and improved upon. Presently, much of the research is still in development and concrete applications are limited, but this paper will show that the current work already shows great potential.

GANs belong to the family of generative models in machine learning ML, generative models are processes that can generate new data instances, more formally, given a observable variable X and a target variable Y, a generative model is a statistical model of the joint probability distribution on P(X|Y). Generally, the process involves discovering patterns in the input data and learning them in a way that the model can then generate new samples that retain characteristics of the original dataset.

### 2.1   The GAN framework

The original GAN framework estimates generative models through an adversarial process, where two models (usually neural networks) are trained in parallel as represented in figure 2 and described below:

There is a generative model, the Generator (G) that captures the data distribution and generates new data. The second model is a classifier, the Discriminator (D) which estimates the probability that a sample came from the training data and not from G. The training process for the Generator is to maximize the probability of its output being misclassified by the Discriminator.

The Generator is responsible for the generation of data, and the Discriminator has the task of assessing the quality of the generated data and providing feedback to the Generator. These neural networks are optimised under game-theoretic conditions: The Generator is optimised to generate data to fool the Discriminator, and the Discriminator is optimised to detect the source of the input, namely the Generator or the real data set.

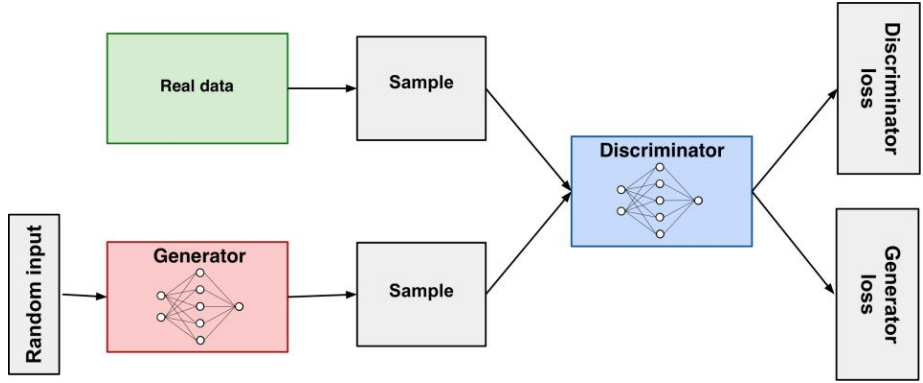

Figure 2: The originally proposed architecture for Generative Adversarial Networks

### 2.1.1 The Discriminator

The Discriminator is a classifier. It receives Real and Synthetic (from G) data and attempts to distinguish them. It can use several different network architectures, depending on the type of data being classified. The Discriminator network connects to two loss functions that are used in different parts of training. After classifying real/synthetic data, the Discriminator loss penalizes it for misclassifications, and its weights are updated via **backpropagation** from its loss through the network.

### 2.1.2 The Generator

The Generator network G uses feedback from the Discriminator D to learn to generate synthetic data that ideally resembles the original data in key aspects. Its goal is for the created data to be classified as real by the Discriminator.

The network receives a random input, some type of noise, from which it generates some output, this output is then evaluated by the Discriminator and results in a Generator loss, which then penalizes the Generator for not deceiving the Discriminator. By introducing noise, a GAN can potentially produce a wide variety of outputs by sampling from different places in the target distribution. Usually noise is introduced by sampling from the uniform distribution.

### 2.1.3 Loss function

The GAN training process uses loss functions that measure the distance between the distributions of the generated and real data to assess their similarity. There are many proposed methods to solve this challenge. In the original "vanilla" GAN, a so-called minimax loss was introduced.

The formulation of the minimax loss is derived from the cross entropy between the real and generated distributions by the JS divergence when the Discriminator is optimal. For the Generator, minimizing the loss is equivalent to minimizing $\log(1 - D(G(z)))$ since it can't directly affect the $\log D(x)$ term in the function. The Jensen-Shannon - JS - divergence measures the similarity between two probability distributions, it is based on the entropy of a discrete random variable being a measurement of the amount of information required on average to describe that variable.

In the original framework, the Generator and Discriminator losses come from a single measure of distance between probability distributions. The two terms are updated in an alternating fashion, depending on which network is being trained.

$$\min_{G} \max_{D} V(D, G) = \mathbb{E}_x[\log D(x)] + \mathbb{E}_z[\log(1 - D(G(z)))]$$

• $D(\mathbf{x})$ is the Discriminator's estimate of the probability that real data instance **x** is real.
• $E_{\mathbf{x}}$ is the expected value over all real data instances.
• $G(\mathbf{z})$ is the Generator's output when given noise **z**.
• $D(G(\mathbf{z}))$ is the Discriminator's estimate of the probability that a fake instance is real.
• $E_{\mathbf{z}}$ is the expected value over all random inputs to the Generator (in effect, the expected value over all generated fake instances $G(\mathbf{z})$).

### 2.1.4 Optimizers

Optimizers update the model in response to the output of the loss function. In essence, they have control over the learning process of a neural network by finding the values of parameters such that a loss function is at its lowest. The learning rate is a key hyperparameter that scales the gradient and sets the speed at which the model is updated.

Most models use a gradient descent-based optimizer. Gradient descent is the direction of steepest descent of a function, these algorithms are used to find the local minimum of differentiable function by small iterations.

Adaptive moment estimation (Adam) is an optimization algorithm used to update network weights iteratively, it is an extension to stochastic gradient descent and has seen broad adoption in deep learning applications from computer vision to natural language processing. Adam works by storing both the exponentially decaying average of past squared gradients and exponentially decaying average of past gradients. There are a few types of optimizers being used in GAN literature, but Adam is currently among the most popular choices.

### 2.1.5 Training

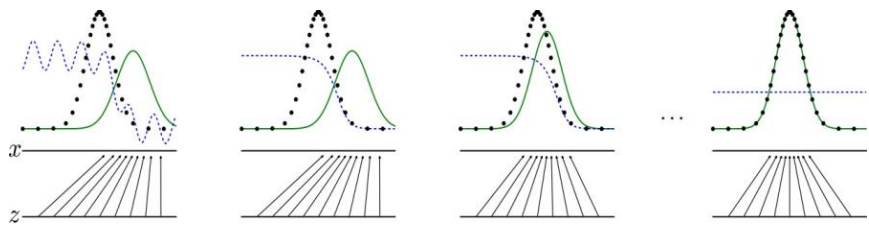

Figure 3: Theoretical evolution of GAN training[6], as the Generative distribution (green line) approaches the real data distribution, the Discriminative distribution (dotted blue line) will be unable to distinguish them and stabilizes at $D(\mathbf{x})$ = ½

The Discriminator (D) and Generator (G) networks are trained separately, by different alternating processes.

**Generator - G**
Sample from random noise z -> new sample is produced in G -> D classifies the sample as "Real" or "Fake" -> Loss calculated from D classification -> Backpropagate through Discriminator and Generator to obtain gradients -> Gradients used to change G weights

**Discriminator - D**
D classifies real data and G fake data -> Discriminator loss penalizes G for misclassifying real and fake instances -> D updates its weights through backpropagation from the Discriminator loss through the network D.

**Adversarial training**
Following the description above, both models are competing against each other in a way called adversarial in game theory, and they are playing a zero-sum game. This means that when the Discriminator successfully identifies a sample, it is rewarded or no update is done to the model parameters, whereas the Generator is penalized with large changes to model parameters. From the other perspective, when the Generator tricks the Discriminator, it is rewarded, or no update is done to its parameters, but the Discriminator is penalized, and its model parameters are changed.

**Using the Discriminator to train the Generator**
When training a neural net, weights are altered to reduce the error or loss of the output. Generative Adversarial Networks are more complex since the Generator is not directly connected to its loss function. It is the Discriminator that produces the output that will affect the Generator (Generator loss). Backpropagation then adjusts each weight by calculating the weight's impact on the output.

The impact of a Generator weight depends on the impact of the Discriminator weights it feeds into. Backpropagation starts at the output and flows back through the Discriminator into the Generator. In an optimal case, where both the Discriminator and Generator evolve at the same pace, the Generator would end up generating samples indistinguishable from those drawn from real data, in a process shown in 3.

### 2.1.6 Challenges

Although capable of generating very accurate synthetic samples[7], GANs are also known[8] to be hard to train. Training two networks simultaneously means that when the parameters of one model are updated, the optimization problem changes. This creates a dynamic system that is harder to control. Non convergence is a common issue in GAN training. Deep models are usually trained using an optimization algorithm that looks for the lowest point of a loss function, but in a two player scenario, instead of reaching an equilibrium, the gradients may conflict and never converge, thus missing the optimum minima. In other words, if the Generator gets too good too fast, it may fool the Discriminator and stop getting meaningful feedback, which in turn will make the Generator train on bad feedback, leading to a collapse in output quality.

**Mode collapse** is a failure mode of GANs that happens due to a deficiency in training. It can happen when the Generator maps several noise input-values to the same output region, or when the Generator ignores a region of the target data distribution. This means that if the Generator gets stuck in a local minimum generating limited samples, the Discriminator will eventually learn to differentiate the Generator's fakes, ending the learning process and leading to an undiversified output. This is a problem because in generative modelling, the goal is not only to create realistic looking samples, but also to be able to produce a wider variety of samples. Several adaptations of the original model try different adjustments to mitigate these problems.

The ongoing GAN research has shown that when the Discriminator gets too good, the training of the Generator can fail. The reason is that an optimal Discriminator doesn't provide sufficient feedback for the Generator to properly learn. This is called the **vanishing gradient** problem, when the gradient gets so small that in backpropagation it does not change the weight values of the Generator's initial layers, so their learning can get very slow and eventually come to a halt.

### 2.1.7 Lack of a proper evaluation metric

Another issue concerning the development of GANs, is how to evaluate their training accuracy. Since GANs have been first developed around image generation, the most used evaluation metric is currently the Fréchet Inception Distance (FID). It is based on the Fréchet Distance, a measurement that compares the statistics of two multivariate normal distributions – mean and covariance matrix - to quantify how far apart the distributions are from each other. It uses the features extracted from the imageNet dataset by the Inception v3 network. Thus, limiting this metric to image generating GANs.

### 2.1.8 What is being done

To solve the main training issues, vanishing gradient and mode collapse, research has developed into two main approaches: The proposal of new network architectures and the use of different loss functions. Recent research[8] shows that the performance of GANs is related to the network and batch sizes, indicating that well designed architectures have critical effect on output quality. The redesign of loss functions, including regularization and normalization has yielded improvements on training stability. It is important to note that improvements are targeted at specific applications, so presently there is not one unique fits-all solution.

## 3 The evolution of GANs in finance - overview and state of research

The possibility of recreating complex distributions via GANs has captured the attention of the quantitative finance researchers, although GANs most notorious application lies in computer vision and image generation, there is some promising research on the field of data-driven modelling in finance, especially its application to time series analysis and generation.

There is currently a steady development undergoing in the broad field of finance, with several applications being researched for generative adversarial networks. Some of the principal developments, also portrayed in table 1, are in market prediction[9], tuning of trading models[10], portfolio management[11] and optimization [12], synthetic data generation[3] and diverse types of fraud detection[13].

When modelling financial time series, there are several models used and developed over the years, such as the before mentioned ARCH/GARCH models, Black Scholes (1973) and Heston (1993). Until the advent of machine learning, made possible by the ever-increasing available computational power, development of financial models has been slow. Purely data-driven modelling via neural networks and machine learning is a growing sub-field of research[14]. The application of GANs has the potential to improve the modelling of complex and unknown statistical dynamics present in financial data.

## 3.1 About the need for synthetic data

Among the different applications of GANs being researched in finance, the topic of generation of synthetic datasets deserves awareness. Synthetic data is important because it can be tailor-made for specific uses or conditions where real data may be lacking or unavailable. This can be useful in numerous cases such as:

• Privacy and compliance rules may severely limit data availability and application[15].
• Data is often required in product testing environments and is often limited or unavailable to testers[16].
• Machine learning requires large amounts of training data, such data can be expensive and scarce[17].

The growth of data-driven processes and the new possibilities provided by this kind of analytics and modelling created a higher demand for data and data scientists in finance. Financial datasets are often very regulated, which limits their use in developing new products and processes. As described by JP Morgans' AI Research[18], anonymization is unreliable, and encryption can cripple the use cases of data. Statistically consistent synthetic financial data can solve most of these limitations, producing high-volume artificial test data also greatly improves scalability and cooperative work with usually sensitive or limited datasets.

Table 1: GANs in finance research

| Field | Application | Method |
|---|---|---|
| Time Series Forecasting | Market Prediction | GAN-FD [9], ST-GAN [19], MTSGAN [20] |
| | Fine-Tuning of trading models | C-GAN [10], MAS-GAN[21] |
| Portfolio Management | Porfolio Optimization | PAGAN[11], GAN-MP[22], DAT-CGAN[23], CorrGAN[12] |
| Time Series Generation | Synthetic time series generation and Finance Data Augmentation | TimeGAN[24], WGAN-GP[25], FIN-GAN[3], Quant GAN[14], RA-GAN[26], CDRAGAN[27], SigCWGAN[28], ST-GAN[19] |
| Fraud Detection | Detection of market manipulation | LSTM-GAN[13] |
| | Detection of Credit Card Fraud | RWGAN[29], LSTM-GAN-2[30] |

## 3.2 Main GANs in financial research

There have been several GAN variants proposed in the literature to improve performance, these can mainly be divided into two types: Architecture and Loss variants. In the Architecture variants, structural changes were made to adapt the GAN to a certain purpose, or to improve overall performance. In Loss variants, different approaches to Loss functions try to improve stability and performance while training, often trying to solve the issue of non-convergence. Modifications have been made to tailor each network to its specific goal and used data type. Overall, the main topic of GAN research is and remains centred around image generation and computer vision. Even so, based on the continuing output of time series and finance applied models it is clear that GANs are helping to expand the field of research. There have been some milestone papers which will be discussed in the next segment.

### 3.2.1 FIN-GAN - 2019

FIN-GAN[3] is a proposed application of the original GAN with the goal of generating synthetic time series that replicate the main stylized facts of financial time series. There are no relevant structural changes to the first proposed framework. For the architecture of D and G, three structures were proposed, 1- MLP (multi layer perceptron) architecture with four layers of fully connected neural networks with the hyperbolic tangent activation function. 2- CNNs (convolutional neural networks) architecture with six convolutional layers and the leaky ReLU activation function. 3- MLP-CNN which combines CNN and MLP by the element-wise multiplication of their outputs.

The algorithm remains generally unchanged from the original proposition, with minimax loss and the Adam optimizer for updating D and G parameters. The main findings were that changes on the Generator's neural network architecture have a greater effect on the quality of output then changes on the Discriminator. The use of batch normalization (rescaling hidden vectors to keep mean and variance consistent during training) showed a large fluctuation in the generated time series and a strong autocorrelation. An indication that this common tool for image generation in deep learning is not ideal for modelling of financial time series. The output is set to a large value because the length of the time-series should be long enough to reliably calculate the statistical properties such as the probability distribution and the correlation functions.

**FIN-GAN applied to finance**
FIN-GAN can generate financial time series that are deemed realistic, due to the presence of the major stylized facts, showing the characteristics that are intrinsic to time series found in the real world. The approach marks an important point for GANs in finance, since it is a model that learns the properties of data without requiring explicit assumptions or mathematical formulations, something that stochastic process and agent-based modelling cannot do without non-trivial assumptions.

### 3.2.2 Conditional GAN (cGAN) - 2019

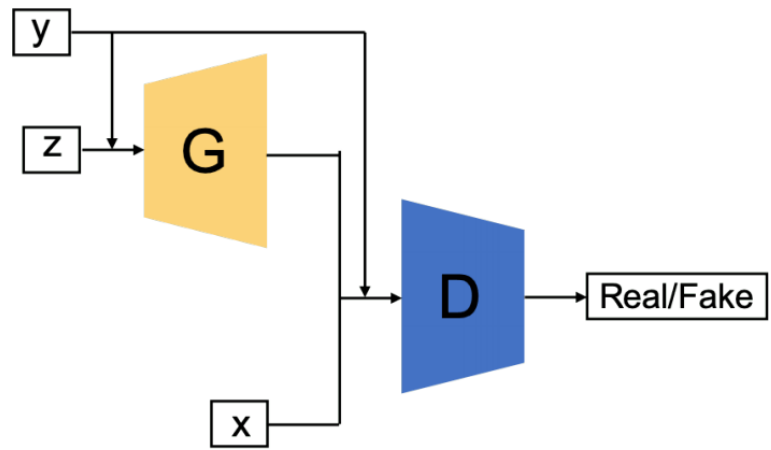

Figure 4: Conditional-GAN architecture[8]

The Conditional GAN (cGAN) was first introduced in 2015[31], and later adapted to finance by Koshiyama [10] 2019. This GAN's architecture has been structurally extended to a conditional model (figure 4) where the Generator and Discriminator are conditioned on some extra information y. This auxiliary information y could take the form of class labels or other data. Which is then fed into both D and G as additional input layers, in a process called conditioning.

The added input layer of one-hot-encoded labels guides the Generator to produce specific outputs. This level of control does not exist in the originally proposed GAN architecture. With the benefit from the additional encoded information, cGAN can also be trained on multimodal datasets that contain labelled data.

Like the original GAN, the Discriminator and Generator networks are MLP's where the Generator output layer activation is a linear function and the Discriminator output layer activation is a sigmoid function. The loss function has an added conditional element for both D and G, and they both are optimized via stochastic gradient descent.

**Conditional GAN applied to finance**

Conditioning information in finance: added vector y can represent a current/expected market condition, appropriate for modelling sequential data such as time series. It also enables the construction of "what-if" scenarios, commonly used for stress tests and other scenario-based models. The implemented cGAN has shown some new tools to improve trading strategies, specifically for finding optimal hyperparameters (fine-tuning) of trading algorithms, and combination of trading strategies, where trading models are trained on samples generated by the cGAN, and their outputs combined to get a final result.

Compared to the classical method of bootstrapping (resampling), the cGAN can generate more diverse training and testing sets. There is also the capability of drawing samples targeting stress events and an added layer of anonymization to the dataset, not achievable on other shuffling and resampling techniques.

### 3.2.3 WGAN-GP - 2019

The first Wasserstein GAN (WGAN[32]) proposed a new cost function using Wasserstein distance as an alternative to the original GAN Loss function, which is based on the KL and JS distances. The Wasserstein distance quantifies the distance between two probability distributions, it is also called Earth Mover's distance because it can be interpreted as the minimum energy cost of reshaping one probability distribution into another via stepwise (discrete) increments, analogous to transporting blocks of dirt.

This modification attempts to solve the problem of vanishing gradient and mode collapse which happened with BCE Loss GANs: Often when the Discriminator started to improve faster than the Generator, it would start outputting more extreme values, giving less meaningful feedback to the Generator, stopping the learning process. With W-Loss the Discriminator is replaced by a Critic. The Critic replaces the sigmoid with a linear activation function, this way the output is not limited between 0 and 1 and the cost function continues to grow, regardless of how far apart the distributions are. This way, the gradient won't approach zero, making the GAN less prone to vanishing gradient problem, and consequently, mode collapse.

There is a condition however, when using W-Loss, the Critic NN needs to be 1-Lipschitz continuous. Meaning that the norm of the gradient should be at most one at every point, this condition assures that the W-Loss function is continuous (differentiable) and that its growth is stable during training, this makes the underlying earth mover distance valid. To enforce this condition WGAN uses weight clipping, where the weights of the Critic are forced to take values between a fixed interval. After the weights are updated during gradient descent, any weights outside of the desired interval are clipped, so values that are too high or too low are set to some fixed maximum value. Forcing the weights to a limited range of values could limit the critics ability to learn and ultimately for the GAN to perform. This causes the need for a lot of hyperparameter tuning to adjust the model for training.

In WGAN-GP[33], another way to enforce the 1-L continuity was introduced, Gradient penalty or GP, which is a much softer way to enforce this continuity. With GP a regularization term is added to the loss function, it penalizes the critic when its gradient norm is higher than 1. The regularization term is achieved by getting a gradient on the interpolation between samples from the real and generated distributions.

**WGAN-GP applied to finance**

The finance application of WGAN-GP[25] for the one-dimensional case proposes a Wasserstein's GAN with Gradient Penalty with 1-dimensional convolutional networks to work on time series data. It compared the relevant statistics of generated versus original time series and proposed taking a rolling window before feeding the original series to the model, aiming to induce more variability scenarios. In the results, the generated data was largely able to replicate the stylized facts of the S&P 500 while also retaining some visual similarity with visible volatility clusters.

### 3.2.4 Corr-GAN - 2019

CorrGAN[12] is a novel approach to generate empirical correlation matrices estimated on asset returns. This GAN is based on the DCGAN[34] architecture.

DCGAN is a milestone GAN, being the first to apply deconvolutional neural networks in the Generator (figure 5). Some critical changes to the architecture of DCGAN compared to original GAN, which enabled higher resolution modelling and stabilized training. It also introduced the learning of hierarchies of representations in natural images. Deconvolutions are also known as transposed convolutions, they work by swapping the forward and the backward passes of a convolution. There are several key aspects of DCGAN:

• The DCGAN structure replaces pooling layers with strided convolutions for Discriminator and fractional-strided convolutions for Generators.

• Batch normalization in both the Discriminator and the Generator, improving location of generated real samples that center at zero.

• A ReLU activation is used in Generator for all layers except the output, which uses Tanh, while LeakyReLU activation is used in the Discriminator for all layers. The LeakyReLU activation prevents the network from being stuck in a "dying state" situation as the Generator receives gradients from the Discriminator.

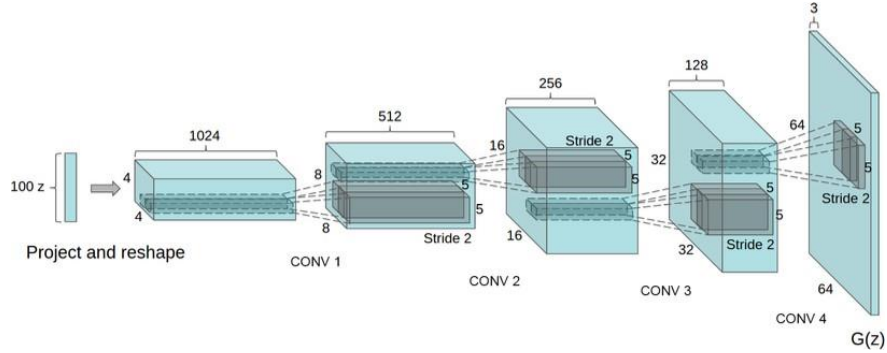

Figure 5: Generator layout of DCGAN[34]

**CorrGAN applied to finance**

The DCGAN architecture was originally developed to improve the quality and resolution of synthetic images. The CorrGAN paper showed an implementation of DCGAN on empirical correlation matrices estimated on S&P 500 returns. The process showed convincing results, although the generated matrices are not exactly correlation matrices (their diagonal is very close but note equal to 1), the major stylized facts were reproduced in the synthetic data. In short, the main stylized facts of financial correlation matrices are the distribution of pairwise correlations being positively shifted, the first eigenvector having positive entries and the corresponding Minimum Spanning Tree being scale free.

Practical financial applications of the CorrGAN framework could range from improving trading strategies to risk and portfolio stress testing. Some suggested applications include improving Monte Carlo backtesting and stress testing portfolios by conditioning on market regime variables of different macroeconomic scenarios. The researchers developed a website[35] to test if people can visually distinguish real from fake correlation matrices, currently they are indistinguishable to the naked eye.

### 3.2.5 QuantGAN - 2020

Introduced in 2019, and one of the first solid works applying GANs to financial time series generation, QuantGAN[14] consists of a GAN variation which utilizes temporal convolutional networks (TCNs) aiming at capturing long-range dependencies like volatility clusters. The objective was to approximate a realistic asset price simulator by using a neural network, data-driven concept.

The TCN are convolutional frameworks that provide a unified approach to capture low and high level information in a single model. TCNs consist of a causal 1D fully convolutional network architecture, where an output depends only on past sequence elements. Another key characteristic are more dilated convolutions. Dilation refers to the distance between elements of the input sequence that are used to compute one element of the output sequence, this means that instead of increasing the size of the kernel/filter, it introduces empty spaces for more coverage, increasing the receptive field and giving a broader view, with more context information about the input.

Empirical results suggest that TCNs are better at capturing long-range dependencies in time series than other convolutional methods. TCNs have the advantage of having more controllable gradients compared to Recurrent Neural Networks, simplifying the optimization process.

**QuantGAN applied to finance**

QuantGAN uses TCNs as Discriminator and Generator networks. It was trained on nine years of S&P 500 data and used the Wasserstein distance as a distributional evaluation metric. As dependence scores, it used the ACF and leverage effect scores to assess accuracy based on the stylized facts. The result was that QuantGAN generated returns closely matched the real returns, this was corroborated by the sharp drops in the ACF and the negative correlation between squared and non-squared log returns at short time lags.

Finally, the authors evaluate that the proposed architecture generated competitive results, which can be used to approximate financial time series. They point to the lack of a unified metric to quantify the goodness of fit of two datasets, but overall the findings were a solid step in developing data-driven models in finance.

### 3.2.6  MAS-GAN - 2021

This new framework, developed by JP Morgan's AI Research team in a yet to be published paper[21], is a two-step method for multi-agent market-simulator calibration. It uses a GAN Discriminator to calibrate a market simulator constructed using an agent-based approach, first a calibration objective is learned with a GAN Discriminator, then the learnt calibration objective is used to simulate parameter optimization. It is the first method to use a GAN trained Discriminator as an objective function for multi-agent system optimization.

This work builds on SAGAN[36], an architectural variant developed in 2018 that introduces a self-attention layer to a Convolutional GAN. Using Convolutional Neural Networks CNNs, the Self-Attention method was designed to create a balance between efficiency and long-range dependencies (large receptive fields) by using the attention mechanism, well known from the Natural Language Processing (NLP) field of research. In a self-attention layer visualized in 6, the feature map is passed through three 1x1 convolutions, called Query, Key and Value, where the vectors generated by Query and Key multiply to create a vector of probabilities that decides how much of Value to expose to the next layer. In other words, the Q x V multiplications' output is passed through a softmax activation function that creates a so-called attention map, which is multiplied by the Value vector to create the self-attention feature map. This layer is complementary to regular convolutions, aides the network in capturing fine features from an input source.

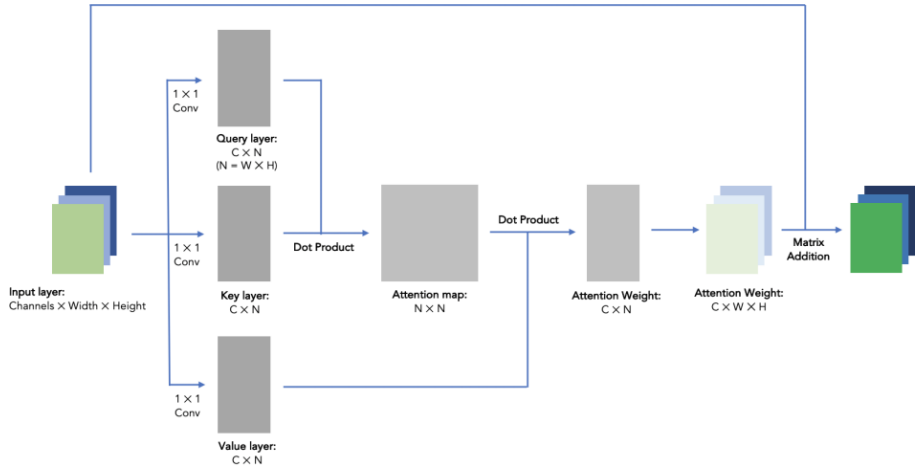

Figure 6: Self-Attention Module[37]

**MAS-GAN applied to finance**
The MAS-GAN model consists of a SAGAN trained on one-minute mid-price returns and cumulative one-minute traded volumes of thirty stocks traded on NASDAQ in June 2019. After the training process is done, the Discriminator is used to optimize a Discrete event model that simulates a Limit Order Book via three distinct agent behaviours. The objective is to find the most realistic agent configuration through calibration, namely, to find which configurations on a rectangular grid that can produce the simulation closest to the historical dataset. The method has shown good calibration performance for multi-market simulations, based on stylized facts and the Kolmogorov-Smirnov test. Further research into understanding behavioural explanations of the market is suggested, but not present in the paper.

### 3.3  Current GAN applications in the financial industry

As shown in the previous section, The topic of Generative Adversarial Nets is progression in different finance applications, but the extent of its use in practice by companies is presently limited or not fully disclosed. This section provides a short view of the publicly known implementations of GANs.

### 3.3.1 Generation of Synthetic data by JP Morgan AI Research

With the goal of advancing AI research and development in financial services, JP Morgan's AI Research department has a branch dedicated to generating synthetic datasets. These datasets can be requested by other research groups and comprise of 1- customer related datasets for Anti Money Laundering models, 2- customer journey events, lower level client-bank interaction dataset, 3- market execution data: limit order book data describing matches of buy and sell orders of financial instruments at a public stock exchange. 4- payment data for fraud detection: several transaction types with legitimate and abnormal activities to improve detection.

The research department proposes a framework[18] for ideal representation and transference of synthetic data. The framework suggests: 1- Privacy preserving, the specific data and context where privacy needs to be enforced. 2- Human readability, the data and its associated generative models must be readily interpretable by regulators and other agents for the sake of transparency. 3- Compactness, the representation of synthetic data should be compact and reconstructible, it should require little technical know how as to improve synthesizing it in different environments.

All the specified data is produced in-house, there is no specific information on the generating process, but there is a paper under review by JP Morgan researchers where the MAS-GAN[21] is proposed for multi-agent simulation, where the Generator is used to calibrate an agent based model as described earlier in this paper.

### 3.3.2 The Digital Sandbox Pilot by the FCA

The Financial Conduct Authority is a financial regulatory body from the United Kingdom. To boost innovation, they created an environment for testing of financial models, products, and services. The Digital Sandbox[38] provides an integrated, collaborative development environment for testing and scaling projects, aiming to reproduce real scenarios and perform stress tests. The initiative is in pilot stage and has already had a test run with 28 groups presenting solutions on the topics of: access to finance for SMEs, improving the financial resilience of vulnerable consumers and fraud/scam detection.

The whole Digital Sandbox heavily relies on synthetic data, as real is under strict obligations in the UK. As described in the first report[16] synthetic financial data was commissioned to leading data scientists from industry and academia. There was a two-group effort, where one group would create the synthetic data and another who could define typologies and behaviours these data was expected to have. The main approaches to this task used GANs and ABMs and the whole process was benchmarked by the Alan Turing Institute. Access to synthetic data was ranked as the most important feature of the sandbox, however, not all data was considered useful, showing that there is demand for high quality synthetic data.

The main takeaways from the first pilot were: A digital testing environment is in high demand, particularly by startups. It accelerated product development for the participating firms. The access to good synthetic data was considered extremely valuable by all participants. A second digital sandbox is planned with an expanded testing system. This promising new platform has shown potential and could be implemented in other countries if it further succeeds in the United Kingdom.

### 3.3.3 Fraud detection by American Express AI Labs

Fraud losses, majorly from wire transfers and credit/debit cards caused an estimated loss of 16.9 billion USD to banks, merchants, and customers[39]. Companies use their customer data to train models to predict and prevent fraud. A known issue of fraud detection on real datasets is the class imbalance: Often a dataset can represent a biased sample from reality and will inevitably lead to faulty models. Financial services multinational American Express Co. has its AI lab looking for solutions to this issue by generating synthetic data to improve their fraud detection models. Amex researchers published a paper[27] where a hybrid of Conditional and Deep Regret Analytic GANs was proposed to generate synthetic datasets.

Three tabular datasets with internal company data were used to recreate statistically similar samples. The generated data was evaluated by comparing characteristics of the real and generated data distributions, and also by the internally developed tool DataQC[40], which uses well known methods to look for anomalies in datasets and outputs a unified score of attribute anomaly levels. The generated data showed satisfactory results, but it was found that models trained on synthetic data still performed worse than those trained on real data. The research team states that further research into generating synthetic data is being done, the capacity on which the generated data is being used internally was not disclosed.

## 4 Application of GANs for time series generation

### 4.1 Experimental setup, a black-box approach

To test the performance of GANs from a data-driven perspective, a black-box framework was used to build a proof of concept model. The idea behind the black box approach is to test whether a model would show good performance with minimal user input, avoiding the complex task of tuning the model, this would greatly improve its usability by people that may not be specifically trained in deep learning or GANs, broadening the range of its application. Another reason for this approach is to evaluate the originally proposed GANs on their out-of-the-box performance on financial data, this excludes the task of tuning GANs from the scope of this project. In the experiment, the goal is to test the performance of selected GANs in capturing the structure of financial time series and generate a diverse set of output scenarios.

**Setting and implementation**
The framework[41] was imported to Google Colab, a hosted Jupyter notebook running the programming language python, that provides free access to computing engines, namely Google's cloud-based CPUs and GPUs. This simplifies the testing/reproducibility in collaborative environments and eliminates the need for local GPUs.

The framework consists of 12 different GANs, implemented mirroring their original proposition. The user has a choice of one GAN to be trained at a time. The framework was extended by adding a training section for inputting real time series data, since the original code was designed around image generation, the input had to be reshaped to accept uni-dimensional time series. The added modifications allow for time series to be imported from Yahoo finance, transformed to log-returns, and adjusted to fit the dimensions expected by the model. The data analysis was centred around a series of evaluation plots that were implemented to test the data generated by the model for its similarity to some of the known key statistical characteristics -stylized facts- of financial time series. The models are trained using log returns, expecting outputs with similar characteristics. The models are trained with around 5000 data points and for the sake of comparison generate outputs of the same size.

From the pool of available GANs three were used: DCGAN[34], SAGAN[36], WGAN-GP[33], the choice was made due to the differences in architecture and loss functions among the available options, and their implementation to finance data in the current literature. The used GANs follow the originally proposed structure and parameters, which were already described in the previous section.

• **DCGAN** was selected because it is a milestone GAN, the first work to apply a deconvolutional neural network architecture for the Generator which showed a bump in resolution to the generated images.
• **WGAN-GP** has been shown to be successful in improving training stability by providing an alternative loss function with the use of Wasserstein's loss.
• **SAGAN** is an alternative GAN based on convolutional neural networks, with an added self-attention mechanism that improves learning on long-range dependencies for images, achieving great performance on multi-class image generation.

These GANs were chosen due to their successful past implementation on financial data. However, they were not specifically tuned for time-series generation, and their performance is therefore not comparable to papers with specific time-series implementations.

### 4.1.1 The S&P 500 and synthetic data generation

Instead of using prices, most studies use returns of assets. The main reason is that prices are usually non-stationary (their mean and variance change over time). Using returns makes the time series stationary, greatly facilitating the process of statistical modelling. Returns also provide a scale free summary of the investment, further improving its flexibility for analysis.

In the experiment, the dataset used to train the models, consists of a 20 year cut of the S&P 500 index (from 01.01.2001 to 01.01.2021, roughly 5000 data points). The S&P 500 is a weighted stock market index that combines 500 of the largest companies listed on US stock exchanges. It was used as it condenses the behaviour of the american economy over time to a single measurement, and is largely used to test financial time series models.

The raw S&P 500 adjusted close values were transformed to log returns for model training and later evaluation. There was no single measure of goodness of fit used, rather a qualitative analysis over the statistical properties of the generated time series where the selected stylized facts described in the first section of this paper were used as reference for performance, which in this case represents how close the stylized facts of generated returns are from those commonly found in real-world returns.

## 4.2 Statistical analysis of time series

In testing the three selected GANs, the main goal is to test whether the models can successfully capture the structure inherent to financial time series and generate scenarios that can be considered diverse. The main issue is the lack of a unified metric or consensus on how to quantitatively evaluate the synthetic samples. GANs do not provide an explicit representation of the generated probability distribution, making it difficult to estimate their likelihood, thus making the task of evaluating them more difficult.

So, the process of evaluation heavily relies on the key selected stylized facts, for which each model is qualitatively assessed based mostly on evaluation plots. By illustrating these key statistics, it's possible to assess some level of performance of the models. This following subsection contains the data analysis performed on the generated returns, which will give a summary of the statistics mainly concerning the distribution of the synthetic returns as well as some time-dependent properties in order to check for the similarity to the real returns.

### 4.2.1 Analysis of returns

By looking at the log-returns plot from the S&P 500 (figure 7), it is possible to recognize some key patterns. Firstly, the limit in which the returns fluctuate, centered at zero and staying between -10% and 10%. Second, the clusters of periods with high volatility and low volatility are usually bundled together, as asserted by the stylized fact of **volatility clustering**. This pattern is visually identifiable, and it is expected that a good synthetic sample resemble the original data by repeating this patterns of clusters.

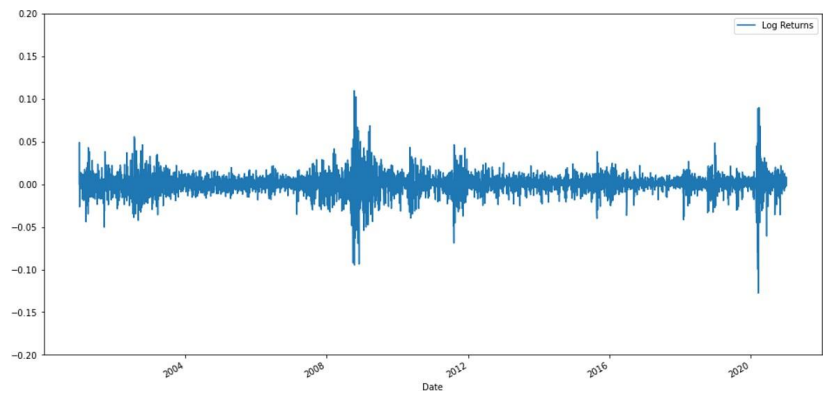

Figure 7: Log-returns of SP500 - It is possible to visually identify periods of high and low volatility

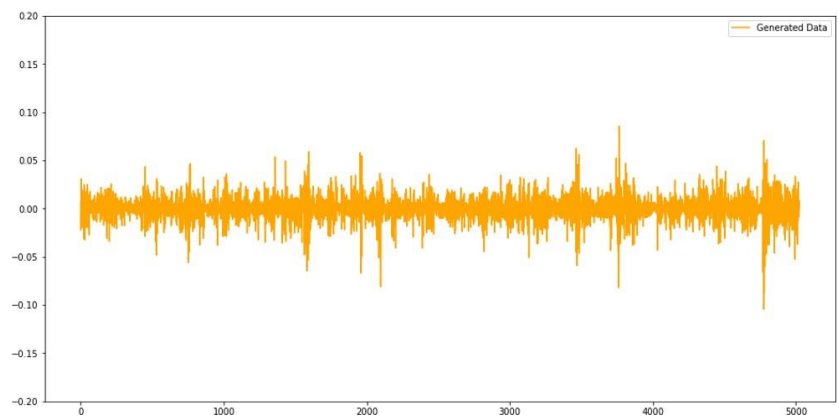

Figure 8: Returns generated by WGAN-GP

Looking at the returns generated by WGAN-GP (figure 8), it is possible to see some similarities to the real returns from S&P 500. The generated returns fluctuate in the same range as the real returns, also centered at zero, and there is some clustering going on. The issue is that the models struggled to recreate realistic clusters, as the output from WGAN-GP is clearly more homogeneous then the real data, the clusters are less pronounced.

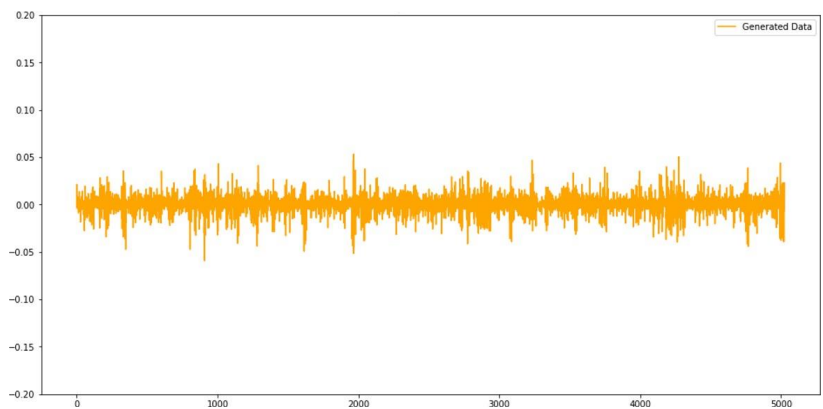

Figure 9: Returns generated by DCGAN

In the returns generated by DCGAN (figure 9), the result was less convincing than the previous models. The centering around zero is present, but there are almost no identifiable clusters, and the synthetic returns did not reproduce the larger values giving the plot a homogeneous aspect that only loosely resembles a real set of log-returns.

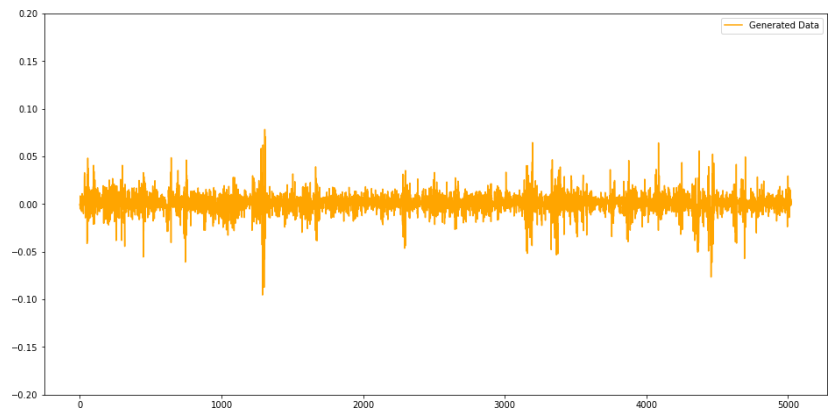

Figure 10: Returns generated by SAGAN

In the SAGAN generated returns (figure 10), the clusters are identifiable, the range and centering is correct. The issue is that the model failed to recreate realistic clusters. The synthetic clusters are too thin, diverging from their real counterparts where more activity is present around large movements.

#### 4.2.2 Probability distribution, Kurtosis and Skewness

Here the generated and real returns were displayed to analyze their distribution and its characteristic skewness and kurtosis measurements. For a GAN to be qualified as good, it is expected that the synthetic returns distribution resemble the shape of the normal distribution (**aggregational gaussianity**) and the values for skewness and kurtosis should be close to those from real returns.

•*Skewness* is a measure of the asymmetry of the probability distribution of a real-valued random variable about its mean. Here it portrays the **gain/loss asymmetry** of log returns, where it is expected that the skewness be negative, since it is more common to have large downwards movements in stock prices then the contrary.

•*Kurtosis* is a measure that defines how much the tails of a given distribution differ from the tails of a normal distribution. It identifies whether there are extreme values in the tails of a distribution. For returns, large values for kurtosis (**heavy tails**) are expected, since some low probability events have a large impact on the distribution. This behaviour does not happen in a normal distribution, so returns are expected to have the rough shape of a normal distribution, but with much heavier tails.

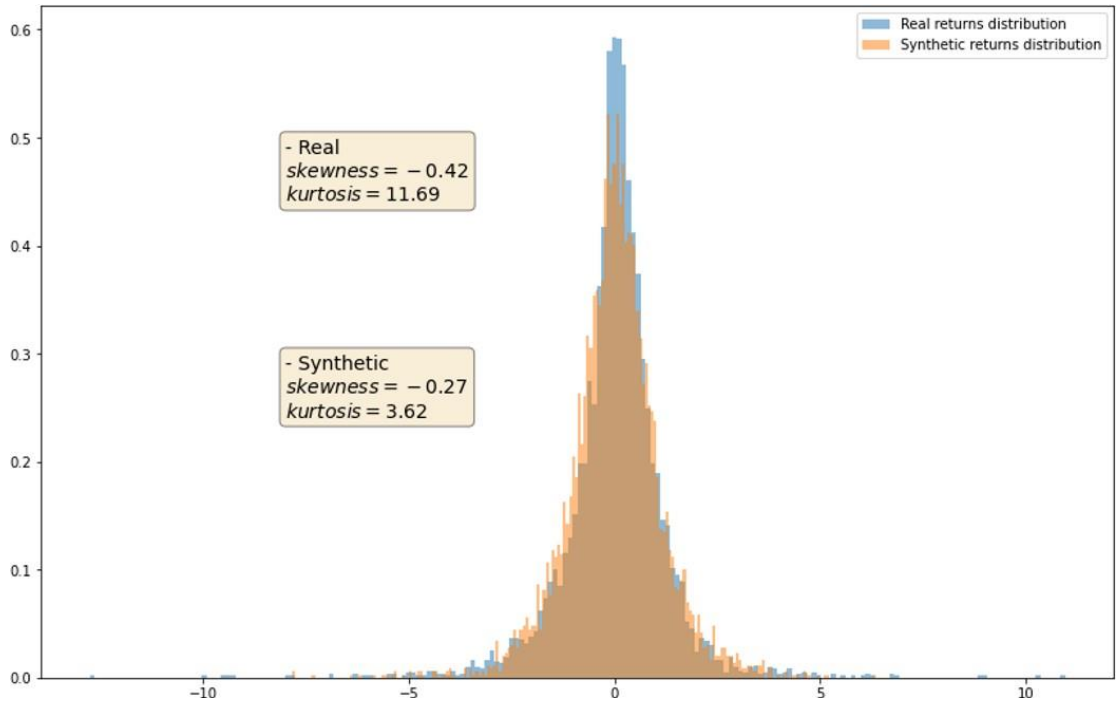

Figure 11: PDF - WGAN-GP

In WGAN-GP's distribution (figure 11), the skewness is negative and close to the value from the real samples. Kurtosis is also positive, but the overall distributions do not share the exact same format.

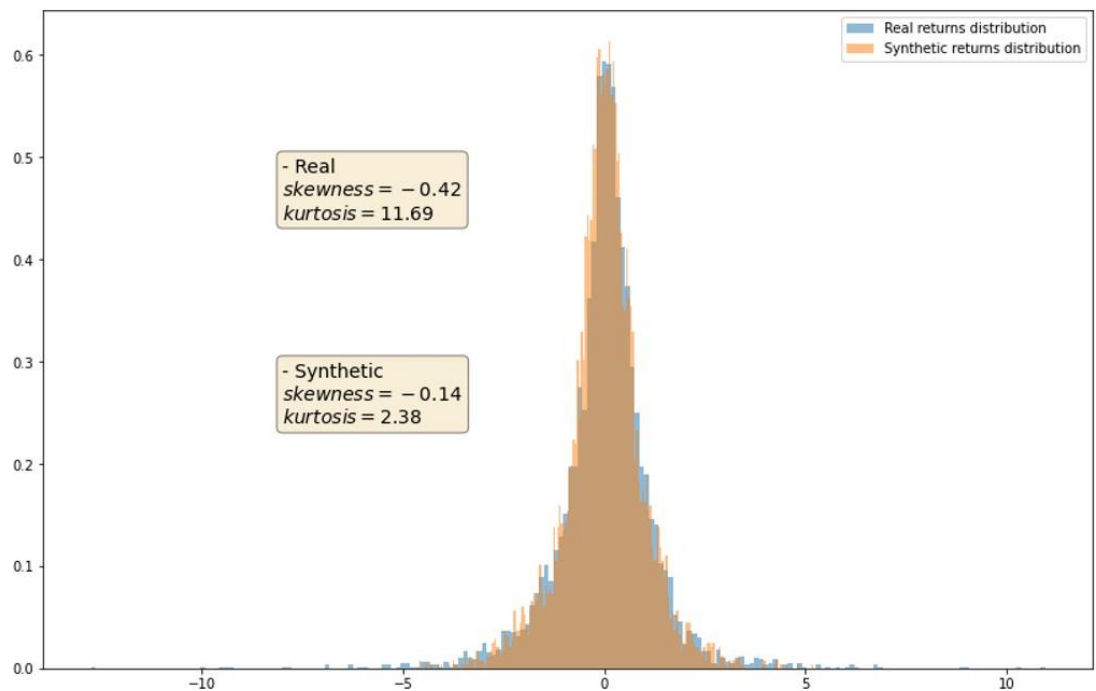

Figure 12: PDF - DCGAN

DCGAN (figure 12) and SAGAN (figure 13) show a good fit, getting close to the distribution observed in the S&P 500 dataset. The are negative and relatively close, and kurtosis is postive, although here SAGAN displays heavier tails, closer to real data. In summary, all three models produce a good skewness but fail to recreate the intensity of heavy tails present in the real distribution.

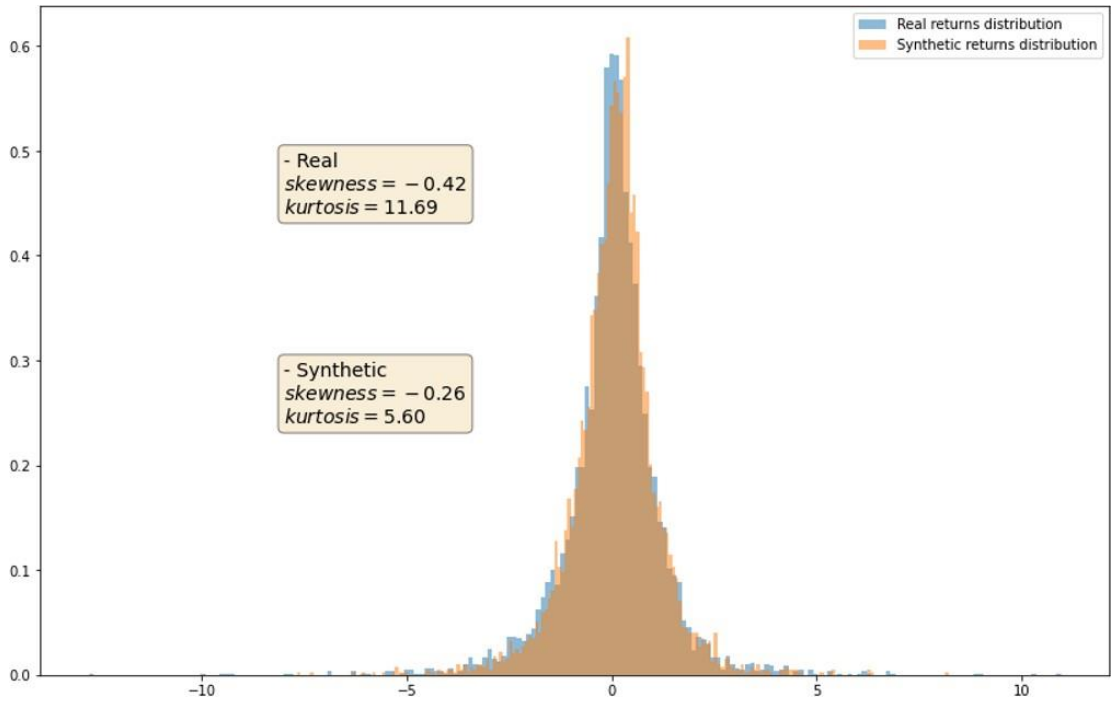

Figure 13: PDF - SAGAN

### 4.2.3 Autocorrelation

The auto-correlation plots (ACF) show the similarity between observations as a function of time lags between them, for financial returns the autocorrelation it is expected to be very low, given the stylized fact of **linear unpredictability**.

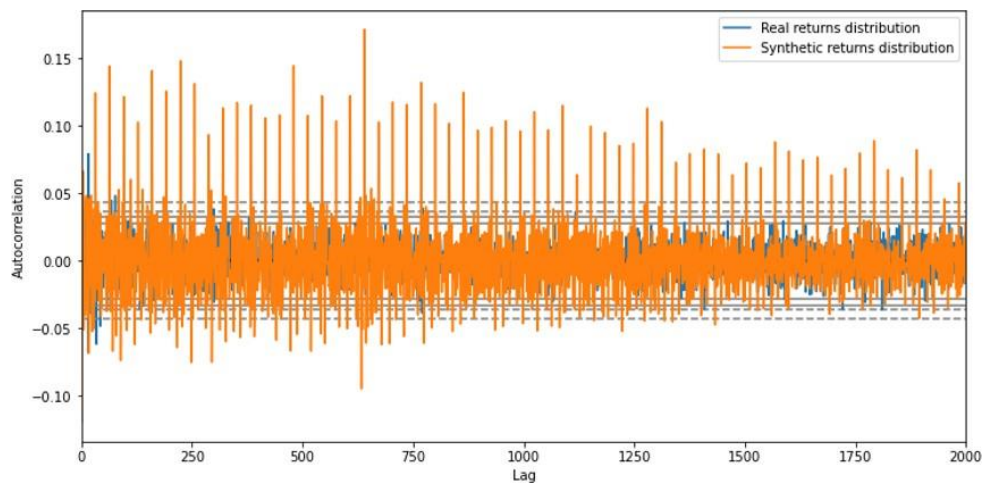

Figure 14: ACF - WGAN-GP

Here its clear that WGAN-GP (figure 14) fails to reproduce the expected linear unpredictability, as the auto-correlation is above the threshold.

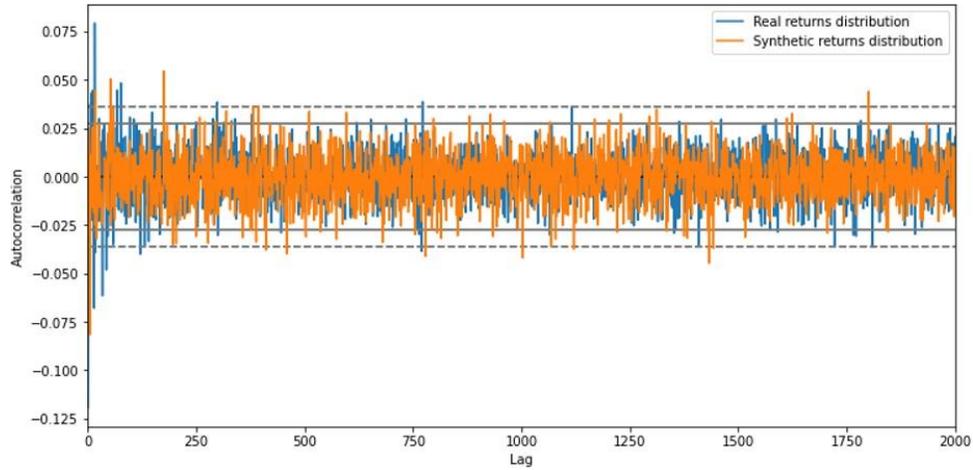

Figure 15: ACF - DCGAN

DCGAN's synthetic samples (figure 15) display a neglectful auto-correlation, precisely mimicking the behaviour of real returns.

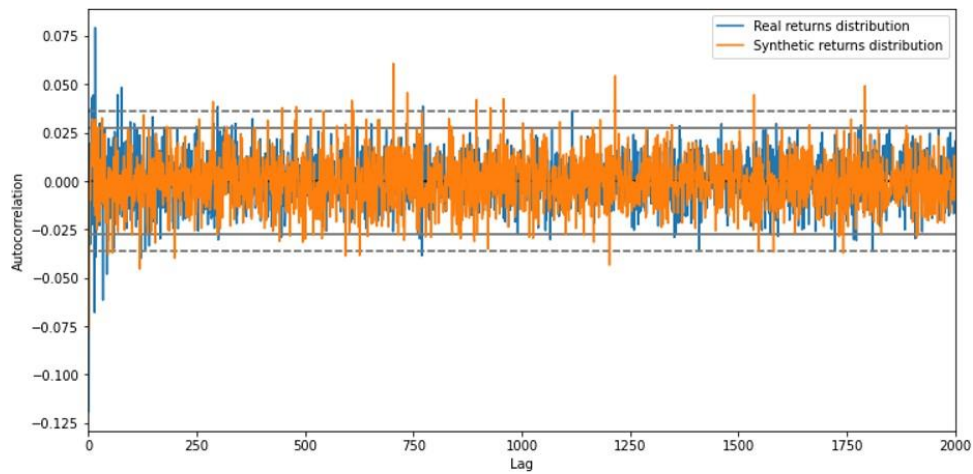

Figure 16: ACF - SAGAN

The auto-correlations in SAGAN's generated returns (figure 16) peek a little bit above the threshold, but overall its a good fit and will be considered satisfactory for this experiment.

### 4.2.4   Transformed prices

Transforming the generated returns back to prices, for the sake of visualization, showed a moderate amount of diversity to each GAN, with WGAN-GP (figure 17) and DCGAN (figure 18) showing realistic looking series, inside the scale of the SP500 index, and SAGAN (figure 19) reproducing a good shape but completely missing a realistic price interval.

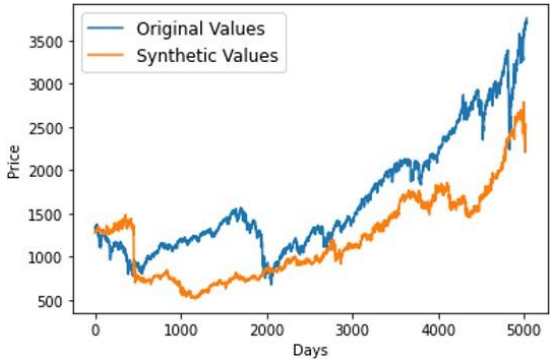

Figure 17: Prices - WGAN-GP

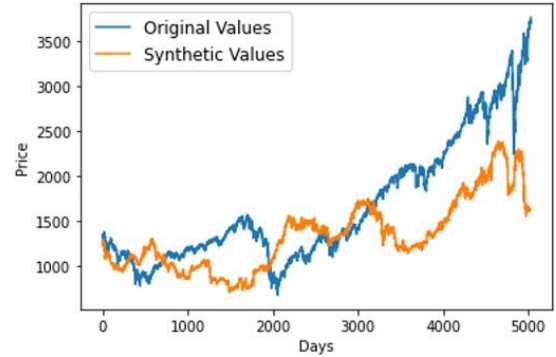

Figure 18: Prices - DCGAN

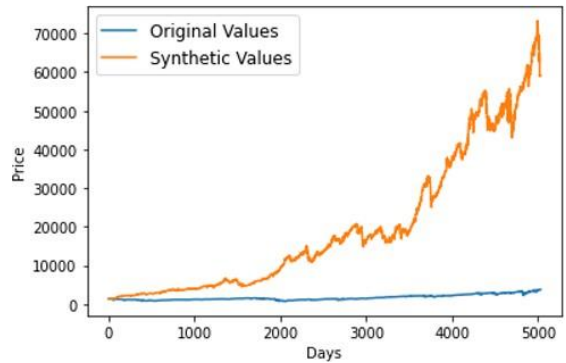

Figure 19: Prices - SAGAN

## 4.3 Results

Although referred to as qualitative statistics, the stylized facts and their proxy statistics can give a good perspective on the performance of the models trained in the experiment section. It was clearly possible to distinguish a best out of the box model among those tested. DCGAN performed surprisingly well considering that its original purpose was image generation it was able to capture mostly all the tested statistical properties of financial time series. Considering that the model was not tuned for this specific case, it shows that with further research into tuning and architectures, GANs have a strong potential for time series modelling.

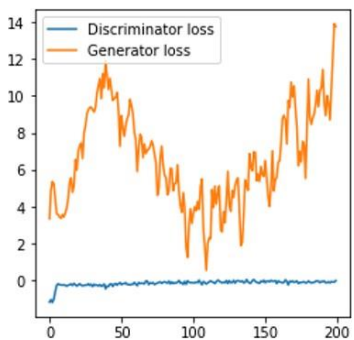

Figure 20: Loss - WGAN-GP

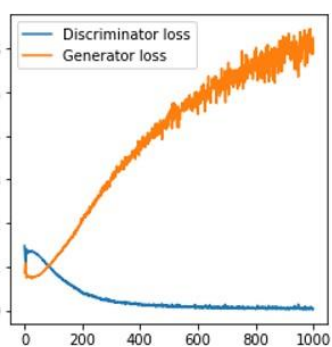

Figure 21: Loss - DCGAN

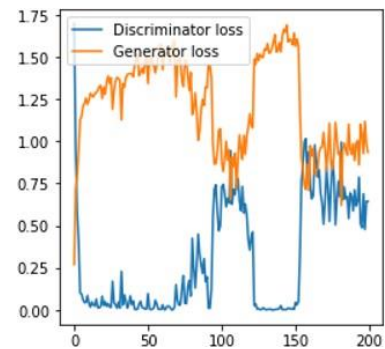

Figure 22: Loss - SAGAN

It was possible to observe some form of mode collapse in all models, it is not possible to quantify this failure mode, but visually each batch of generated data had some visual similarities in the shape of returns and prices, the produced samples did not achieve a large variety of scenarios. The issue of loss convergence was also present during training, where SAGAN notably failed to settle even after 1000 epochs, and in WGAN-GP / DCGAN the losses appeared to settle. It is important to note that this does not mean that a model that converges performs better, or that its best output is at the point of convergence. In fact, for WGAN-GP (figure 20) and SAGAN (figure 22), the model performed better before the losses appeared to settle, and only DCGAN (figure 21) performed best at the max setting of 1000 epochs.

The training was limited to 1000 epochs at fixed learning rates. With more computational power available, longer training epochs could potentially lead to better results. For a next step in these tests, more parameter tuning and better computational power would be advised to potentially get better results.

## 5 Conclusions - Outlook for the future of GANs in finance

### 5.1 Analysis of results

This paper had the goal of presenting an overview of Generative Adversarial Networks in finance, describing what they are, why they are being used in finance, followed by some key developments in research and practical applications. To test the versatility of different frameworks out of the box, the objective was to test three notorious image generation GANs on time series.

The importance and wide application of synthetic data in finance was shown as well as some of the challenges faced in the task of generating it. An overview of the finance GANs in literature was presented, along with an insight on some milestone papers in this field. The active research shows that generating synthetic financial data is achievable and viable with the use of GANs, other novel uses such as the calibration of trading models has also shown positive results. The research has refined the training of GANs with finance data, but it is still a complex task, and further stabilizing training will remain an active research topic. A remaining issue is the lack of a unified quantitative metric to assess the performance on generated financial data. Most literature relies on the reproduction of stylized facts, which is a sound metric, but a more precise score would greatly benefit the development of these models, as previously acknowledged in the literature review.

Although being a novel approach, GANs are making their way to the financial industry. Some applications were shown, specially the use of synthetic data generated by adversarial models is gaining traction. The usefulness of good synthetic data is gaining traction, since it opens the possibility of modelling information that would otherwise be regulated for privacy reasons or too scarce for deep models. It was shown that GANs can generate consistent data for various purposes, including retail banking and market data.

As a proof of concept, three known GANs were tested on time series data to assess their out-of-the-box performance. There were no clear expectations regarding the results, but the models performed well. Using selected stylized facts as qualitative metrics, the synthetic generated returns were close to their real counterparts. Considering that the models where tested out of the box, with no hyperparameter tuning except for epoch length, the results are quite positive and attest for the power of generative adversarial networks. Since the main problem with GANs is in training and financial time series generation is still an ongoing research topic, the various complex aspects of GAN training were out of the scope of the tested models.

### 5.2 Recommendations

Overall, GANs have a large potential in solving the specific need for synthetic financial data, along with other diverse modelling tasks. Being still a novelty in finance, they are more present in research than in practice, but their potential can increase the capabilities of data-driven deep models, reaching areas where these methods previously couldn't be applied due to limitations in data. The development of better evaluation metrics for sample comparison is a key aspect for the further advance in the field, since providing easier tools would make it easier for use by less qualified personnel. Refinements to improve stability and reliability in the training process are in active research and there are still some improvements to be made before a state-of-the-art model is presented. Taking all of this into account, Generative Adversarial Networks offer a good prospect to the modelling toolbox, adding new possibilities to the challenging world of quantitative finance.